\documentclass[twocolumn,prl,superscriptaddress]{revtex4-1}
\usepackage{amsfonts}
\usepackage{amsmath}
\usepackage{amssymb}
\usepackage{color}
\usepackage{graphicx}
\usepackage{bm}
\usepackage{esint}
\usepackage{ulem}
\usepackage{xcolor}

\usepackage{soul}
\usepackage{cancel}

\begin{document}

\title{Photon-assisted charge-parity jumps in a superconducting qubit}

\author{M. Houzet}
\affiliation{Univ.~Grenoble Alpes, CEA, IRIG-Pheliqs, F-38000 Grenoble, France}
\author{K. Serniak}
\affiliation{Departments of Physics and Applied Physics, Yale University, New Haven, CT 06520, USA}
\author{G. Catelani}
\affiliation{JARA Institute for Quantum Information (PGI-11), Forschungszentrum J\"ulich, 52425 J\"ulich, Germany.}
\author{M. H. Devoret}
\affiliation{Departments of Physics and Applied Physics, Yale University, New Haven, CT 06520, USA}
\author{L. I. Glazman}
\affiliation{Departments of Physics and Applied Physics, Yale University, New Haven, CT 06520, USA}

\begin{abstract}
We evaluate the rates of energy and phase relaxation of a superconducting qubit caused by stray photons with energy exceeding the threshold for breaking a Cooper pair. All channels of relaxation within this mechanism are associated with the change in the charge parity of the qubit, enabling the separation of the photon-assisted processes from other contributions to the relaxation rates. Among the signatures of the new mechanism is the same order of rates of the transitions in which a qubit looses or gains energy. 
\end{abstract}

\maketitle

{\bf Introduction --}
One of the most promising approaches toward developing robust hardware for quantum information processing is based on superconducting devices in the circuit-QED setting~\cite{Blais2004,Koch2007,Paik2011,Devoret2013,Girvin2011}. Extending the energy-relaxation time $T_1$ and phase-coherence time $T_\varphi$ of Josephson-junction-based qubits is one of the main goals of this field. Both of these timescales can be limited by the presence of nonequilibrium quasiparticles in the superconducting electrodes~\cite{Martinis2009,Catelani2011,Catelani2011b,Lenander2011,Catelani2012,Gustavsson2016}. In popular transmon qubits, quasiparticle-induced transitions between different states of the qubit are accompanied by the transfer of a single unpaired electron (charge $e$) across the Josephson junction: a signature which can be detected experimentally by monitoring the ``charge parity'' of the device~\cite{Sun2012,Riste2013,Bal2015,Serniak2018,Serniak2019}. This has the effect of changing the transmon offset charge by $e$, so we henceforth refer to these as $e$-jumps. Recently, experiments have directly correlated qubit transitions with $e$-jumps~\cite{Riste2013,Serniak2018}, indicating that these processes can contribute significantly to $T_1$ in state-of-the-art transmons.

In this work, we develop the theory of $e$-jumps caused by the absorption of stray photons with energy $\hbar\omega>2\Delta$ capable of breaking Cooper pairs (here $\Delta$ is the BCS energy gap). The electric field associated with the photon is concentrated at the junction, thus increasing the probability of breaking a pair in its vicinity and transferring an electron across the junction. The $e$-jump may or may not be accompanied by the change in the state of the qubit, so here we evaluate the rates of all transitions that change charge parity. Consequently, we uncover a subtle quasiparticle interference effect that becomes prominent for photons with energy close to $2\Delta$. The photon-assisted $e$-jump rates differ drastically from those caused by  resident steady-state quasiparticles~\cite{Martinis2009,Catelani2011,Catelani2011b,Catelani2014}, thus providing a clear fingerprint of decoherence by photon absorption. The theory explains several recent  experimental findings~\cite{Serniak2018}.

{\bf Photon-assisted $e$-jump rates --}
The role of quasiparticles in an elementary superconducting qubit is captured by the electronic Hamiltonian
\begin{equation}
\label{eq:H}
\hat H_\text{el}=\hat H_\varphi+\hat H_\text{qp}+\hat H_T.
\end{equation}
The first term here describes the quantum dynamics of the superconducting phase difference across a Josephson junction, 
\begin{equation}
\label{eq:Hphi}
\hat H_\varphi=4E_C(\hat N-{n_g})^2-{E_J}\cos\hat \varphi+\frac 12 E_L(\hat \varphi-2\pi\Phi_e/\Phi_0)^2,
\end{equation}
where $\hat \varphi$ and $\hat N=-i d/d\hat \varphi$ are canonically conjugate quantum variables describing the superconducting phase difference and the number of Cooper pairs that tunneled across the junction, respectively; $E_J$ and $4E_C$ are the Josephson and charging energies associated with these two variables; $n_g$ is a dimensionless gate voltage that accounts for offset charges. The inductive shunt of a fluxonium~\cite{Manucharyan2009} is described by the last term in Eq.~(\ref{eq:Hphi}); its presence allows one to use an external magnetic flux $\Phi_e$ to tune the qubit levels ($\Phi_0$ is the superconducting flux quantum). A transmon does not have a shunt, $E_L=0$. Our theory is equally applicable to any device. The eigenstates of Eq.~\eqref{eq:Hphi} are the qubit states $|n\rangle$ with energy $E_n$. The second term in Eq.~\eqref{eq:H} describes quasiparticles residing in the superconducting leads,
\begin{equation}
\hat H_\text{qp}=\sum_{k\sigma}\varepsilon_k \hat \alpha_{k\sigma}^\dagger \hat  \alpha_{k\sigma}+\sum_{p\sigma} \varepsilon_p \hat\gamma_{p\sigma}^\dagger \hat \gamma_{p\sigma}.
\end{equation}
Here $\hat\alpha_{k\sigma}$ is a fermionic  annihilation operator for a Bogoliubov quasiparticle in orbital state $k$ and with spin $\sigma$ in one of the leads, $\hat\gamma_{p\sigma}$ plays a similar role for a quasiparticle in the other lead  ($\sigma=\pm$ for up and down spins); the quasiparticle energy $\varepsilon_k=\sqrt{\xi_k^2+\Delta^2}$ is expressed in terms of the normal-state electron energy $\xi_k$ measured from the Fermi level.  
Finally, the third term in Eq.~\eqref{eq:H} describes electron tunneling across the junction,
\begin{equation}
\label{eq:HT}
\hat H_T=\sum_{kp\sigma} \left[t e^{i\hat\varphi/2} \hat a^\dagger_{k\sigma}  \hat c_{p\sigma} +\mathrm{H.c.}\right]+ E_J\cos{\hat\varphi};
\end{equation}
it accounts for the coupling between $\hat\varphi$ and quasiparticle degrees of freedom. Here the tunnel matrix element $t$ is related to ${E_J}$ through the Ambegaokar-Baratoff relation, $ E_J=g_T\Delta/4$, where $g_T=4\pi^2 \nu_0^2|t|^2$ is the conductance of the junction in the normal state, in units of  $e^2/\pi \hbar$, and $\nu_0$ is the normal density of states per spin. The operator $\hat a_{k\sigma}=u_k\hat\alpha_{k\sigma}+\sigma v_k  \hat\alpha^\dagger_{k\bar\sigma}$ (with $\bar\sigma=-\sigma$) annihilates an electron in one of the leads, and $u_k,v_k=\sqrt{(1 \pm \xi_k/\varepsilon_k)/2}$ are BCS coherence factors (relations for the electron annihilation operator $\hat c_{p\sigma}$ in the other lead are similar). The last term in Eq.~\eqref{eq:HT} is included to avoid double-counting the Josephson energy term appearing in Eq.~\eqref{eq:Hphi}~\cite{Catelani2011b}.

\begin{figure}[tb]
\includegraphics[width=0.8\columnwidth]{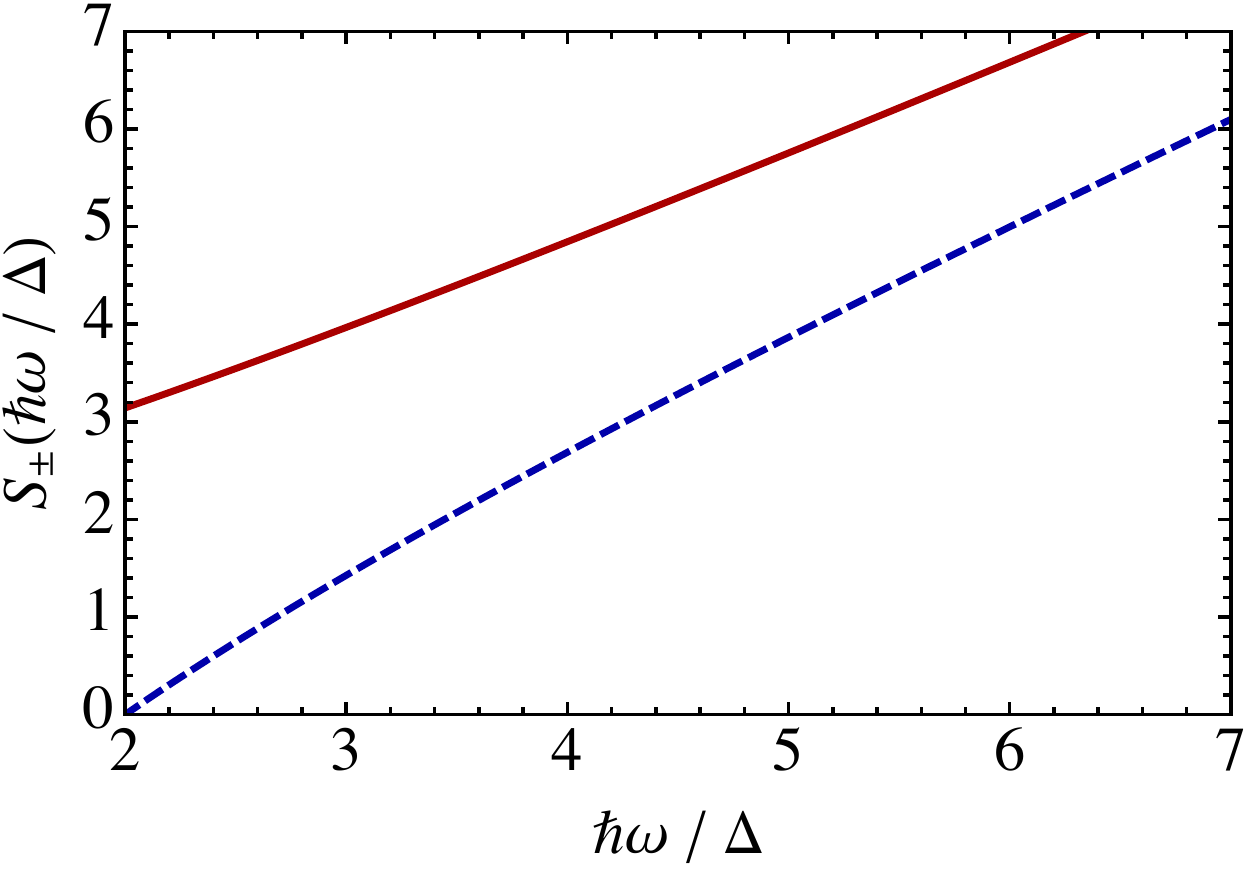}
\caption{\label{Fig:rate} 
Quasiparticle structure factors $S_+$ {(solid line)} and $S_-$ {(dashed line)}~\cite{note:structure-factor-functions} as a function of energy at $\hbar\omega\geq 2\Delta$.
}
\end{figure}

The coupling of the electronic degrees of freedom to an electromagnetic mode in the cavity is described by the Hamiltonian
\begin{equation}
\label{eq:Htot}
\hat H=\hat H_\text{cav}+\hat H_\text{el}\,,\quad \text\quad
\hat H_\text{cav}=\hbar \omega_\nu \hat b_\nu^\dagger \hat b_\nu,
\end{equation}
provided that we make the substitution 
\begin{equation}
\label{eq:rule}
\hat \varphi \to \hat \varphi +\phi_\nu(\hat b_\nu+\hat b^\dagger_\nu)\quad\text{with}\quad\phi_\nu=2e{\cal U}_\nu /(\hbar \omega_\nu)
\end{equation}
in $\hat H_\text{el}$. Here $\hat b_\nu$ is the bosonic annihilation operator for a cavity mode $\nu$ with frequency $\omega_\nu$ and operator of the electric field $\hat {\cal E}({\bf r})=-i{\cal E}_\nu({\bf r})(\hat b_\nu-\hat b^\dagger_\nu)$. The ``zero-point fluctuation" of the phase, $\phi_\nu$, and voltage drop, ${\cal U}_\nu$, across the Josephson junction are proportional to the electric field ${\cal E}_\nu({\bf r})$; for definiteness, we relate ${\cal U}_\nu$ to the field value at the junction, ${\cal U}_\nu=d_\nu{\cal E}_\nu(0)$. In general, the effective length $d_\nu$ depends not only on the specific geometry of the qubit, but also on the frequency $\omega_\nu$. Inserting the substitution rule \eqref{eq:rule} into Eq.~\eqref{eq:HT} and accounting for the weakness of coupling ($\phi_\nu\ll 1$), we express the Hamiltonian of the quasiparticle-photon-qubit interaction as
\begin{eqnarray}
\delta \hat H_T&=&\frac{i\phi_\nu}2(\hat b_\nu+\hat b^\dagger_\nu) \left(\hat V_1+\hat V_2\right)+\text{H.c.},
\\
\hat V_1&=&t
 {\sum_{kp\sigma} }
 \left(e^{i\hat \varphi/2}u_ku_p+e^{-i\hat \varphi/2} v_kv_p\right) \hat \alpha^\dagger_{k\sigma}  \hat \gamma_{p\sigma},
\nonumber\\
\hat V_2&=&t
\sum_{kp\sigma} 
\sigma \left(e^{i\hat \varphi/2}u_kv_p-e^{-i\hat \varphi/2} v_ku_p\right) \hat \alpha^\dagger_{k\sigma}  \hat \gamma ^\dagger_{p\bar\sigma}.
\nonumber
\end{eqnarray}
Treating $\delta \hat H_T$ as a perturbation to $\hat H_0=\hat H_\varphi+\hat H_\text{qp}+\hat H_\text{cav}$, and assuming a vanishing occupation of the quasiparticle states, so to neglect $\hat V_1$, we can use the Fermi's Golden Rule to evaluate the rate for absorbing a cavity photon while changing the qubit state from $n$ to $m$,
\begin{eqnarray}
\label{eq:FGR}
\Gamma_{nm}=\frac{2\pi}{\hbar}\left(\frac{\phi_\nu}2\right)^2
\sum_{kp\sigma}|\langle{\text{vac}},m| {\hat\alpha_{k\sigma}\hat\gamma_{p\bar\sigma}}\hat V_2|{\text{vac}},n\rangle|^2
\quad\quad
\\
\times
\delta(\hbar\omega_\nu+E_n-E_m-\varepsilon_k-\varepsilon_p),
\nonumber
\end{eqnarray}
where $|{\text{vac}},n\rangle=|{\text{vac}}\rangle\otimes|n\rangle$ is the product of the BCS ground state and the qubit state. Evaluating the sums in Eq.~\eqref{eq:FGR}, we can express the $e$-jump rates~\eqref{eq:FGR} as
\begin{eqnarray}
\label{eq:rates0}
\Gamma_{nm}=\Gamma_\nu\left[|\langle n|\cos\frac{\hat\varphi}2|m\rangle|^2S_-\left(\frac{\hbar\omega_\nu+E_n-E_m}\Delta\right)\right.
\quad\\
\quad\left.+|\langle n|\sin\frac{\hat\varphi}2|m\rangle|^2S_+\left(\frac{\hbar\omega_\nu+E_n-E_m}\Delta\right)\right]
\nonumber
\end{eqnarray}
with the common characteristic scale
\begin{equation}
\label{eq:rate-ampl}
\hbar \Gamma_\nu=\frac {2}{\pi }\left(\frac{2e{{\cal U}_\nu}}{\hbar \omega_\nu}\right)^2 {E_J}
\end{equation}
for the photon absorption. Properties of the superconducting quasiparticles are accounted for by the dimensionless structure factor functions ($x=\hbar\omega/\Delta$), {see Fig.~\ref{Fig:rate},}
\begin{equation}
\label{eq:SFF}
S_\pm(x) {=}\int _1^\infty dy\int _1^\infty dy'\frac{yy'\pm1}{\sqrt{y^2-1}\sqrt{y'^2-1}}\delta(x-y-y'),
\end{equation}
with the following asymptotes:
\begin{equation}
\begin{array}{lr}
S_\pm(x)=0,&\quad x<2,\\
S_+(x)= {\pi[1+(x-2)/4]},&\quad x-2\ll 2,\\
S_-(x)={(\pi/2)(x-2)},&\quad x-2\ll 2,\\
S_\pm(x)\approx  {x},&\quad x\gg 2.
\end{array}
\label{eq:asymptoteS}
\end{equation}
Their prefactors inside the brackets of Eq.~\eqref{eq:rates0} are matrix elements for the transitions between qubit states. While these matrix elements also enter into $e$-jump rates due to residual quasiparticles, the structure factor functions are different~\cite{Catelani2014}.

At $E_J\gg E_C$, Eq.~\eqref{eq:Hphi} describes a weakly anharmonic oscillator whose phase displays small quantum fluctuations around the classical phase $\varphi_0$.  At finite $E_L$ it may be tuned away from zero by an external flux $\Phi_e$, and found as the solution of equation
$E_J\sin\varphi_0+E_L(\varphi_0-2\pi \Phi_e/\Phi_0)=0$,
which yields the minimum of (classical) energy. In the harmonic approximation, Eq.~\eqref{eq:Hphi} reduces to
$\hat H'_\varphi=4E_C(\hat N-n_g)^2+ {\tilde E_J}(\hat\varphi-\varphi_0)^2/2$ with $\tilde E_J=E_J\cos\varphi_0+E_L$. The weak anharmonicity singles out the ground and excited states of the qubit. Retaining only the lowest-order correction in $(E_C/\tilde E_J)^{1/2}$, one obtains $\hbar \omega_{01}\approx \sqrt{8\tilde E_JE_C}{-E_C}$ for the corresponding transition frequency. Evaluation of the qubit matrix elements in Eq.~(\ref{eq:rates0}) within the leading order~\cite{footnote2} in $(E_C/\tilde E_J)^{1/2}$ yields
\begin{widetext}
\begin{subequations}
\label{eq:rates}
\begin{eqnarray}
\Gamma_{00}&=&\Gamma_{11}= \Gamma_\nu\left[\frac{1+\cos\varphi_0}{2} S_-\left(\frac{\hbar\omega_\nu}\Delta\right)+
\frac{1-\cos\varphi_0}{2} S_+\left(\frac{\hbar\omega_\nu}\Delta\right)\right],\label{eq:rate-gg}\\
\Gamma_{01}&=& \Gamma_\nu\sqrt{\frac{E_C}{8\tilde E_J}} 
\left[\frac{1+\cos\varphi_0}{2} S_+\left(\frac{\hbar\omega_\nu-\hbar\omega_{01}}\Delta\right)+
\frac{1-\cos\varphi_0}{2} S_-\left(\frac{\hbar\omega_\nu-\hbar\omega_{01}}\Delta\right)\right] ,\label{eq:rate-ge}\\
\Gamma_{10}&=& \Gamma_\nu\sqrt{\frac{E_C}{8\tilde E_J}} 
\left[\frac{1+\cos\varphi_0}{2} S_+\left(\frac{\hbar\omega_\nu+\hbar\omega_{01}}\Delta\right)+
\frac{1-\cos\varphi_0}{2} S_-\left(\frac{\hbar\omega_\nu+\hbar\omega_{01}}\Delta\right)\right] .\label{eq:rate-eg}
\end{eqnarray}
\end{subequations}
\end{widetext}
The $\varphi_0$-dependence of the rates \eqref{eq:rates} reveals the interference between quasiparticles crossing the junction in the photon-absorption process. It is reminiscent of the $\cos\varphi$-effect in the dissipative Josephson current~\cite{Barone1982} and flux-dependent fluxonium relaxation rates~\cite{Pop2014}. At $E_L\neq 0$, frequency $\omega_{01}$ is independent of $n_g$, which can be gauged out from $H'_\varphi$. Sensitivity of the qubit energy levels to the gate voltage is useful for separating out the rates of various $e$-jump processes~\cite{Riste2013,Serniak2018}. The $\varphi_0$-dependence of the rates \eqref{eq:rates} may be investigated in a device retaining such sensitivity, {\it e.g.}, in a flux qubit~\cite{Bal2015,Yan2016}.

Let us make several observations. First,
at large frequency, $\hbar \omega_\nu\gg2\Delta$, the transition rates are independent of $\varphi_0$, and we find $\Gamma_{01}/\Gamma_{10}\approx 1$ and
\begin{equation}
\label{eq:ratio-large-freq}
\Gamma_{00}/\Gamma_{10}=({{8\tilde E_J}/{E_C}} )^{1/2}.
\end{equation}
Notably, the rates $\Gamma_{00}$, $\Gamma_{11}$ in which a qubit state does not change are substantially larger than $\Gamma_{01}$ and $\Gamma_{10}$.
Furthermore, at $\varphi_0=0$ {and $\hbar\omega_{01}\ll \Delta$} we find
\begin{equation} 
\label{eq:ratio-01-10}
{1-\hbar\omega_{01}/\Delta<\Gamma_{01}/\Gamma_{10} < 1}
\end{equation}
at any frequency above the threshold, $\hbar\omega_\nu>2\Delta+\hbar\omega_{01}$. Finally, at $\hbar\omega_\nu$ close to the threshold, the large factor in Eq.~(\ref{eq:ratio-large-freq}) is compensated by a small factor $S_-(x)$, see Eq.~(\ref{eq:asymptoteS}), resulting in:
\begin{equation}
\frac{\Gamma_{00}}{\Gamma_{10}}\approx \left(\frac{2\tilde E_J}{E_C}\right)^{\!\!1/2}\!\left(\frac{\hbar\omega_\nu}{\Delta}-2\right)\,\,\,\text{at}\,\,\,  \hbar\omega_\nu- 2\Delta\ll \Delta.
\end{equation}

The characteristic rate $\Gamma_\nu$ of Eq.~\eqref{eq:rate-ampl} depends on the qubit parameter $d_\nu$ and the amplitude of the quantized electric field ${\cal E}_\nu$. To estimate the two,  we notice that, in conventional 3D designs~\cite{Paik2011}, the superconducting circuit is oriented along the shortest direction of a 3D electromagnetic cavity, say with a width $L_z$ along $z$-direction much smaller than the characteristic transverse sizes, $L_x,L_y\gg L_z$. 
Therefore, the electric field at frequencies smaller than $\pi c/L_z$ ($c$ is the light velocity) is expressed in terms of the TE modes,
\begin{equation}
{\bm{\hat {\mathcal E}}=-i\sum_\nu(\hat b_\nu-\hat b^\dagger_\nu)}{\mathcal E}_\nu(x,y)\bm{\hat z},
\end{equation}
and is independent of $z$, apart from the vicinity of the qubit and rf input/output connectors constituting perturbing metallic boundary conditions inside the cavity. Furthermore, ${\cal E}_\nu(x,y)$ is a real solution of the equation
\begin{equation}
\label{eq:wave}
\left[\omega_\nu^2+c^2(\partial_x^2+\partial_y^2)\right]{\cal E}_\nu(x,y)=0
\end{equation}
at frequency $\omega_\nu$ in the transverse $(x,y)$-plane, complemented with the appropriate non-radiative boundary conditions defined by the cavity walls and the above mentioned perturbations in the cavity. Because the  perturbations occupy a tiny fraction of the cavity volume, we may disregard them in the normalization condition to obtain
\begin{equation}
\label{eq:E2av}
\overline{ {\cal E}^2_\nu} ={2\pi \hbar \omega_\nu}/(AL_z).
\end{equation}
Here $\overline{ {\cal E}^2_\nu}=(1/A)\int d^2r {\cal E}^2_\nu(x,y)$ and $A$ is the cavity's transverse area.

Given the presence of perturbations, we expect that the boundary conditions associated with Eq.~\eqref{eq:wave} yield a chaotic behavior for its solutions~\cite{Shigehara1998}. The spacing of eigenfrequencies around a given frequency $\omega_\nu$ is estimated as $\delta\omega=c^2/(A\omega_\nu)$. Then, the amplitude of the electric field at the qubit position will fluctuate from mode to mode. We may use the random matrix theory (RMT)~\cite{Beenakker2009} to describe these fluctuations in a range of frequencies of the order $c/\sqrt{A}$ around a frequency $\omega_\nu$ such that $c/\sqrt{A}\ll\omega_\nu\lesssim \pi c/L_z$, where $ \pi c/L_z$ is the cutoff frequency for the TM modes. Therefore, the amplitude of the electric field at a given position is given by the Porter-Thomas distribution in the orthogonal ensemble,
\begin{equation}
\label{eq:Porter}
P({\cal E}^2_\nu)d{\cal E}_\nu^2=(2\pi {\cal E}^2_\nu \langle {\cal E}^2_\nu\rangle )^{-1/2} \exp(-{\cal E}^2_\nu/ 2\langle {\cal E}_\nu^2\rangle)d{\cal E}^2_\nu,
\end{equation}
where the brackets denote the ensemble average. For the modes $\nu$ that can be described with RMT, the spatial and ensemble averages equal each other, $\langle {\cal E}_\nu^2\rangle=\overline{{\cal E}_\nu^2}$.

The effective length $d_\nu$, which characterizes the coupling between the superconducting circuit and the electric field,  is frequency-independent in a wide frequency range. This range is limited by the requirement that the size of the superconducting circuit is smaller than the wavelength of the photon $\lambda_\nu=2\pi c/\omega_\nu$, while inductance $L_J=\hbar^2/(4e^2E_J)$ of the Josephson junction is high enough to treat it as an open circuit, $L_J\omega_\nu\gg Z_{\rm vac}$ (here $Z_{\rm vac}$ is the vacuum impedance). For a typical design, the frequency of stray photons with $\hbar\omega\sim2\Delta$ falls below the upper limit set by the former condition; the much lower transition frequency $\omega_{01}$ exceeds the lower limit set by the latter condition, as long as $\sqrt{E_C/\tilde E_J}\gg \alpha$ (here $\alpha=e^2/\hbar c$ is the fine structure constant). We may extract the frequency-independent $d_\nu\equiv d$ from the dispersive shift measured at the resonator's principal mode frequency $\omega_r$, which is close to $\omega_{01}$. Indeed, ignoring the role of quasiparticles and projecting the Hamiltonian \eqref{eq:Htot} onto the lower-energy states of the qubit yields
\begin{equation}
\hat H=\hbar\omega_r{\hat b^\dagger_r\hat b_r}+\frac{\hbar\omega_{01}}2\sigma_z+\hbar g({\hat b_r+\hat b^\dagger_r})\sigma_x,
\end{equation}
where 
\begin{equation}
\label{eq:g}
g=\frac 1\hbar(2E_C\tilde E_J^3)^{1/4}\frac{2ed{\cal E}_r}{\hbar\omega_r}
\end{equation}
is the ``vacuum Rabi frequency''~{\cite{footnote1}}, and $\sigma_x,\sigma_z$ are Pauli matrices acting in the two-dimensional space of qubit states.
Combining Eqs.~\eqref{eq:rate-ampl}, \eqref{eq:E2av}, and \eqref{eq:g} then yields 
\begin{equation}
\Gamma_\nu=\frac {4}\pi\frac{ g^2}{\omega_{01}}\frac{ \omega_r}{\omega_\nu}\frac{ E_J}{\tilde E_J}
\left(\frac{{\cal E}_r^2(x,y)}{\overline{ {\cal E}_r^2}}\right)^{-1}
\frac{{\cal E}_\nu^2(x,y)}{\overline{ {\cal E}_\nu^2}},
\label{eq:gammanug}
\end{equation}
where $(x,y)$ is the vicinity of qubit location. Using the standard expression for the principal mode in a rectangular cavity, and assuming that the qubit is positioned near the cavity's center, allows to estimate ${{\cal E}_r^2(x,y)}/{\overline{ {\cal E}_r^2}}\approx 4$. 
The ensemble-averaged value of Eq.~\eqref{eq:gammanug} is then
 \begin{equation}
\label{eq:Gamma0}
\langle \Gamma_\nu\rangle=\frac \Delta{\hbar\omega_\nu}\Gamma_0,\qquad
\Gamma_0=\frac {1}\pi\frac{g^2}{\omega_{01}}\frac{\hbar\omega_r}\Delta\frac{ E_J}{\tilde E_J}.
\end{equation}
In the two-level approximation for the qubit states, $g$ is related to the dispersive shift of the qubit transition frequency, $\chi=g^2/|\omega_r-\omega_{01}|$. Equations~(\ref{eq:rates}) and (\ref{eq:Gamma0}) express the main result of this work. 

{\bf $e$-jumps in transmons --} From now on, we specify the discussion to transmons, such that $\tilde E_J=E_J$ and $\varphi_0=0$. There are two aspects in which the rates of charge-parity transitions caused by photons differ qualitatively from those caused by the quasiparticles resident in the qubit. First, it is the approximately equal rates of transitions accompanied by the qubit energy loss or gain, $\Gamma_{01}\approx\Gamma_{10}$, see Eq.~\eqref{eq:ratio-01-10}. 
To the contrary, the resident quasiparticles mechanism~\cite{Catelani2014} leads to $\Gamma_{01}{\ll}\Gamma_{10}$, even if their energy distribution is out of equilibrium~\cite{Catelani2019}. Second, the ratio $\Gamma_{00}/\Gamma_{10}$ is large, see Eq.~\eqref{eq:ratio-large-freq}. In contrast, the quasiparticle tunneling mechanism yields a parametrically smaller result~\cite{Catelani2014}, differing from Eq.~\eqref{eq:ratio-large-freq} by an additional factor $(\hbar\omega_{01} T_{\rm qp}/\pi\Delta^2)^{1/2}\ll 1$; here $T_{\rm qp}\ll\Delta$ is the effective temperature of quasiparticles.

A single photon with energy $\hbar\omega>2\Delta$ is much more effective in causing decoherence than the residual quasiparticle density in a typical setting. This efficiency is a byproduct of the efficient coupling between the superconducting circuit and the electromagnetic cavity in the transmon design. The quasiparticle mechanism~\cite{Martinis2009} yields $\Gamma^{\rm qp}_{10}=x_{\rm qp}\sqrt{2\Delta\omega_{01}/\pi^2\hbar}$, where $x_\text{qp}=n_{\rm qp}/(2\nu_0\Delta)$ is the quasiparticle density in units of the density of Cooper pairs. We compare the effectiveness of a single photon with that of quasiparticles by equating $\langle\Gamma_{10}\rangle=\Gamma^{\rm qp}_{10}$, and finding the corresponding $x_{\rm qp}^{\rm eff}$,
\begin{equation}
x_{\rm qp}^{\rm eff}={\sqrt{2}}\pi^2\alpha\sqrt{\frac{\hbar\omega_{01}}\Delta}\frac{d^2\lambda_\nu}{AL_z}\,.
\end{equation}
For a typical device~\cite{footnote3}, this yields $x_{\rm qp}^{\rm eff}\sim 5\times 10^{-5}$ much larger than the typical residual density~\cite{Paik2011} of $\lesssim 10^{-6}$.

{\bf Comparison with experiment --}
Photon-assisted $e$-jumps provide a natural explanation for the results of the recent experiment~\cite{Serniak2018}. In~\cite{Serniak2018}, the rates of $e$-jumps accompanied by qubit excitation and relaxation, respectively, were approximately equal each other. This observation is consistent with Eq.~(\ref{eq:ratio-01-10}) and hints at a finite probability $n_\nu$ of finding a high-energy photon in the cavity. Furthermore, we may associate the observed rate of $e$-jumps occurring without the qubit leaving the ground state with the rate $n_\nu\Gamma_{00}$,  while the above-mentioned measured rate of $1\to 0$ transitions is associated with $n_\nu\Gamma_{10}$, cf.~Eqs. (\ref{eq:rate-gg}) and (\ref{eq:rate-eg}) with $\varphi_0=0$. Comparing the ratio of the two with the experimental data, we obtain the relation $\Gamma_{00}/\Gamma_{10}\approx {4.1}$, which we treat as an equation for finding the characteristic photon frequency.
Using the qubit parameters~\cite{footnote1}, we find $\omega_\nu\approx 2.8\Delta/\hbar$. Then, inserting this frequency into the ratio of rates \eqref{eq:rate-ge} and \eqref{eq:rate-eg} yields $\Gamma_{01}/\Gamma_{10}\approx {0.97}$, which is close to the observed ratio between qubit relaxation and excitation rates (accompanied by $e$-jumps).

To assess the individual rates (rather than their ratios), we assume the incoming photons belong to a narrow (compared to $\omega_\nu$) bandwidth around the frequency $\omega_\nu$. We also assume this bandwidth is wide compared to the mean-frequency spacing $\delta\omega$, which allows us substitute $\Gamma_\nu$ in Eqs.~\eqref{eq:rates} by its ensemble-averaged value~\eqref{eq:Gamma0}. [In the opposite case of a narrow frequency bandwidth $\ll\delta\omega$, all rates \eqref{eq:rates} would fluctuate from mode to mode according to the Porter-Thomas distribution~\eqref{eq:Porter}; we note also that frequencies $\hbar\omega_\nu\sim 2\Delta$ for the parameters~\cite{footnote1} are at the margin of validity of the used-above condition $\omega_\nu\lesssim\pi c/L_z$.] Inserting the device parameters~\cite{footnote1,footnote3} into Eq.~\eqref{eq:Gamma0}, we find $\Gamma^{-1}_0 \approx {0.6\,\mu\text{s}}$. The measured $e$-jump rates are much lower. This indicates that the measured rates are actually controlled by the probability for a photon to enter the cavity. The sum of all measured $e$-jump rates then yields the rate with which photons appear in the cavity, $dn_\nu/dt=1/T_P$ with $T_P=77\,\mu$s~\cite{Serniak2018}.  Using this value, qubit-state probabilities $P_0\approx P_1\approx 1/2$, and the estimated-above $\Gamma_0$ in equation
\begin{equation}
\label{eq:rate-photon}
dn_\nu/dt=n_\nu\sum_{n,m=0,1}P_n\langle \Gamma_{nm}\rangle,
\end{equation}
we find the photon occupation factor $n_\nu\approx 10^{-2}$. 

Alternatively, we may consider $e$-jumps caused by photons coming into the cavity from outside not within a narrow frequency band, but a distribution corresponding to a thermal bath. In this case, the magnitude of $e$-jump rates depends on the coupling parameters between the cavity modes and the outside bath, which may depend on the frequency of incoming photons. Neglecting such dependence, the poorly known coupling parameter cancels out in the ratios between  rates. In the following estimates, we consider the effect of external irradiation from a thermal bath at temperature $T_\text{b}$. Assuming $\hbar\omega_{01},T_\text{b}\ll 2\Delta$
we find:
\begin{subequations}
\label{eqs:ratesT}
\begin{eqnarray}
\Gamma_{00}/\Gamma_{10}&\approx&
\sqrt{\frac{2E_J}{E_C}}\frac{T_\text{b}}\Delta \exp({-\hbar\omega_{01}/T_\text{b}}),
\label{eq:ratioT00}\\
\Gamma_{01}/\Gamma_{10}&\approx& \exp({-2\hbar\omega_{01}/T_\text{b}}).
\label{eq:ratioT10}
\end{eqnarray}
\end{subequations}
Let us note a poor agreement of the data~\cite{Serniak2018} with such a uniformly-attenuated thermal photons model.
Indeed, equating Eq.~\eqref{eq:ratioT00} with the corresponding observed ratio, together with device parameters~\cite{footnote1}, yields $T_\text{b}\approx 0.68\Delta$. Inserting this temperature into Eq.~\eqref{eq:ratioT10} then yields ratio $\Gamma_{01}/\Gamma_{10}\approx 0.77$, which is about 30\% lower than the observed one. Therefore, we favor the first explanation, involving photons in a relatively narrow frequency band.

{\bf Conclusion} -- 
We have identified a decoherence channel associated with an event of photon-assisted electron tunneling through a Josephson junction in a superconducting qubit. This process results from breaking of a Cooper pair by a stray photon with energy exceeding $2\Delta$, the electric field of which is concentrated at a high-impedance junction. The qubit transition rates accompanying these photon-assisted $e$-jumps are markedly different from those caused by residual quasiparticles, and are consistent with the measured rates in Ref.~\cite{Serniak2018}, where charge-parity switches were equally likely to excite or relax the transmon. Interestingly, we find that the contribution per high-frequency photon in the cavity to qubit decoherence (through energy relaxation) is similar to that of low-frequency photons (through shot-noise dephasing~\cite{Sears2012}). However, this similarity depends on the particulars of the present implementation of the qubit-cavity system. Unsurprisingly, our results reinforce the importance of protecting superconducting qubits from electromagnetic radiation at all frequencies.

\begin{acknowledgments}
We acknowledge stimulating discussions with Y. Alhassid, S. Diamond, V. Fatemi, S. M. Girvin, M. Hays, and R. J. Schoelkopf. This work is supported by NSF DMR Grant No. 1603243,  ARO grant W911NF-18-1-0212, and by the ANR through Grant No.~ANR-16-CE30-0019. GC acknowledges support by the Alexander von Humboldt Foundation through a Feodor Lynen Research Fellowship and hospitality of Yale Quantum Institute.
\end{acknowledgments}


\begin{thebibliography}{50}
\bibitem{note:structure-factor-functions}
Equation~\eqref{eq:SFF} can be expressed analytically in terms of elliptic integrals,
\begin{equation}
\nonumber
 S_\pm(x)=(x+2){E}\left(\frac{x-2}{x+2}\right)-4\frac{x+1\pm 1}{x+2}{K}\left(\frac{x-2}{x+2}\right).
\end{equation}
\end{thebibliography}
\end{document}